\input{aipcheck}
\documentclass[
    ,final            % use final for the camera ready runs
  ]
  {aipproc}

\layoutstyle{6x9}
\def\beq{\begin{equation}}
\def\eeq{\end{equation}}
\def\beqn{\begin{eqnarray}}
\def\eeqn{\end{eqnarray}}

\begin{document}

\title{Phenomenological Quantum Gravity}

\classification{11.30.Cp, 12.60.-i }
\keywords      {Minimal Length, Deformed Special Relativity, Quantum Gravity}

\author{S.~Hossenfelder}{
  address={Perimeter Institute, 31 Caroline St. N, Waterloo Ontario, N2L 2Y5, Canada}
}

\begin{abstract}
Planck scale physics 
represents a future challenge, located between particle physics and general relativity. 
The Planck scale marks a threshold beyond which the old description of spacetime breaks
down and conceptually new phenomena must appear. In the last years, increased efforts
have been made to examine the phenomenology of quantum gravity, even if the full theory is
still unknown.
\end{abstract}

\maketitle

%%%%%%%%%%%%%%%%%%%%%%%%%%%%%%%%%%%%%%%%%%%%
%% MAINMATTER
%%%%%%%%%%%%%%%%%%%%%%%%%%%%%%%%%%%%%%%%%%%%

%\section{Why}

Quantum gravity is probably the most challenging and fascinating problem of physics in the 21st century. 
The most impressive indicator is the number of people working on it, even though so
far there is no experimental evidence that might guide us from mathematics to physical reality. 
During the last years, the priority in the field has undergone a shift towards the 
investigation of possible observable effects\footnote{For an extensive list of references 
see \cite{Amelino-Camelia:2004hm,Hossenfelder:2006cw} and references therein.}. 
The phenomenology of physics beyond the Standard Model (SM) has been studied using effective models. Such  
top-down inspired bottom-up approaches, which use the little knowledge that we have about quantum gravity, 
might soon become experimentally accessible. These models have in common that they are not a fundamentally and fully consistent
description of nature, but able to capture essential features that allow to make testable 
predictions.

Models with extra dimensions are one such approach that has been extensively studied in the last
decade. In some scenarios, the Planck scale can be lowered to values soon accessible at the LHC.
These models predict a vast number of quantum gravitational effects at the lowered
Planck scale, among them the production of TeV-mass 
black holes and gravitons.  

Another approach which is currently under study is to extend the SM by one of the most 
important and general features of quantum gravity that we know: a minimal 
invariant length scale that acts as a regulator in the ultraviolet \cite{Kempf:1994su,Garay:1994en,Ng:2003jk}. 
Such a minimal length scale 
leads to a generalized uncertainty relation and it requires a deformation of Lorentz-invariance 
which becomes important at high boost parameters. 
Within a scenario with large compatified extra dimensional, also 
the minimal length comes into the reach of experiment and sets
a fundamental limit to short distance physics \cite{Hossenfelder:2004gj}.

%\section{How}
 
A particle with an energy close to the Planck mass, $m_{\rm p}$, is expected to significantly disturb space-time on
a distance scale comparable to its own Compton wavelength and thereby make effects of 
quantum gravity become important. A concentration of energy 
high enough to cause strong curvature will result in significant quantum effects of gravity. Such a concentration
of energy might most intuitively be seen as an interaction process. In that process, the relevant energy is that in the center of mass (com) system, which we will denote with 
$\sqrt{s}$. Note that this is a meaningful concept only for a theory with more than one particle. 
However, it can also be used for a particle propagating in a background field consisting of many 
particles (like e.g. the {\sc CMB}). The scale at which effects of quantum gravity become important 
is when $\sqrt{s}/m_{\rm p} \sim 1$, for small impact parameters $\sqrt{t}\sim 1/b \sim m_{\rm p}$.

\begin{figure}
  \includegraphics[height=.3\textheight]{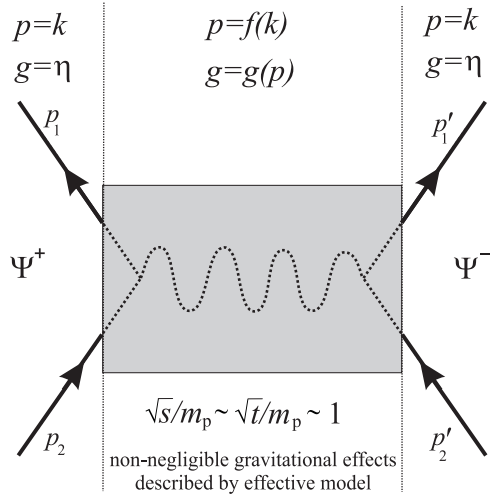}
\caption{In addition to the SM-interaction under investigation, strong gravitational effects accompany
the processes in the collision region. These effects are described in the effective QFT model with
a {\sc MDR} for the exchange particle. Shown is the example of fermion scattering $f^+ f^- \to f^+ f^-$ 
($s$-channel). \label{fig1}}
\end{figure}

The construction of a quantum field theory ({\sc{QFT}}) that self-consistently allows such a minimal 
length makes it necessary 
to carefully retrace all steps of the standard scheme.  
Let us consider the propagation of a particle with wave-vector $k_\mu$, when it comes into a spacetime 
region in which its presence will lead to a com energy close to the Planck scale. The concrete picture we 
want to draw is that of in- and outgoing point particles separated far enough and without noticeable 
gravitational interaction that undergo a strong interaction in an intermediate region which we want to describe in an
effective way\footnote{An effective description as opposed to going beyond the theory of a 
point-like particle.}. We denote the asymptotically free in(out)-going states with $\Psi^+$ ($\Psi^-$), primes are
used for the momenta of the outgoing particles. This is schematically shown in Fig. \ref{fig1}.

In the central collision region, the curvature of spacetime is non-negligible and the scattering process
as described in the {\sc SM} is accompanied by 
gravitational interaction. An effective description of this gravitational interaction 
will modify the propagator, not of the asymptotically free states, but of the particles that
transmit the interaction. The exchange particle has to propagate through a region with strong curvature, and
the  particle's propagation will dominantly be modified by the
energy the particle carries.

The aim is then to examine the additional gravitational 
interaction by means of an concrete model\footnote{Explicit production of real or virtual gravitons, 
or black hole formation are not considered.}. As laid out in \cite{Hossenfelder:2006cw,Hossenfelder:2003jz}
this can be done by either using an energy dependent metric in the intermediate region, or by
assuming a non-linear relation between 
the wave-vector $k$ and the momentum $p=f(k)$, both of which result in a modified dispersion relation ({\sc MDR}), and
make it necessary to adopt a deformation of Lorentz-transformations 
\cite{Amelino-Camelia:2004hm,Magueijo:2002am,Hossenfelder:2005ed}. 

The functional form of the unknown relation $f(k)$ is where knowledge from an underlying theory 
has to enter. The essential property of this relation is that in the high energy limit the wave-vector becomes
asymptotically constant at a finite value, which is the inverse of the minimal length. 
The relation between momentum and wave-vector can be expanded in
a power series where $m_{\rm p}$ sets the scale for the higher order terms.  
 
The wave-vectors coincide with the momenta of the in- or outgoing particles far away from 
the interaction region, where space-time is approximately flat $g \sim \eta$. We denote these 
asymptotic momenta by $p_i$. Putting the interaction in a box and forgetting 
about it, $\sum_i p_i$ is a conserved quantity. The unitary operators of the Poincar\'e group act as
usual on the asymptotically free states. In particular, the whole box is invariant under translations 
$a_\nu$ and the translation operator has the form $\exp(-i a^\nu p_\nu)$ when applied to $\Psi^\pm$.  

In contrast to the asymptotic momenta $p$, the wave-vector $k$ of the particle in 
the interaction region will behave non-trivially because strong gravitational
effects disturb the propagation of the wave. In particular, it will not transform as a standard 
(flat space) Lorentz-vector, and obey the {\sc MDR}.   

Under quantization, the local quantity $k$ will be translated into a partial
derivate. Consequently, under quantization the momentum $p$ results in an  
higher order operator $\delta^\nu = {\rm i} f^\nu (- {\rm i \partial})$. 
 From this one can further define the operator $g^{\mu\nu}(\partial_\alpha ) 
\partial_\mu \partial_\nu = \delta^\nu \partial_\nu$, 
which plays the role of the propagator in the quantized theory. It captures
the distortion of the exchange-particles in the strongly disturbed background.

It is convenient to use the higher order operator $\delta^\nu$ in the setup of a field
theory, instead if having to deal with an explicit infinite sum.
Note, that this sum actually has to be infinite when the relation $p^\nu = f^\nu(k)$ has an 
asymptotic limit as one
would expect for an UV-regulator. Such an asymptotic behavior could never be 
achieved with a finite power-series. 
As an example, the action for a scalar field\footnote{For a discussion of the Dirac-equation, gauge fields, and applications see e.g. 
\cite{Hossenfelder:2003jz,Hossenfelder:2004up}.} takes the form
\beqn
S &=& \int {\rm d}^4 x \sqrt{g }{\cal L} \quad,
\eeqn
with ${\mathcal L} = \delta^\nu \phi~ \partial_\nu \phi$.  

According to the above discussion of the interaction region, one can then examine the properties of the
$S$-matrix. The quantity we measure for ingoing and outgoing states of a collision is typically not 
the wave-length of the particle but its ability to react with other particles. For 
scattering processes, the quantity $k$ therefore is a mere dummy-variable that justifies its existence 
as a useful interpretational device in intermediate steps, where it enters through the propagator, or the {\sc MDR}, respectively. It is in principle possible to calculate 
in $k$-space, however, eventually $k$ can be completely replaced by the physical momentum $p$. It
is important to note that from the construction of the model, $k$-space has finite boundaries, whereas the 
momentum-space is infinite with a squeezed measure at high energies that regulates the 
usually divergent integration (see e.g. \cite{Hossenfelder:2004up}).
 
From the above discussion it is now also apparent that there are two conceptually different ways to include
a minimal length scale into the {\sc QFT}s of the {\sc SM} and to obtain an effective model. The model 
discussed here leaves the transformation of the free single particle unmodified since for such 
a particle there is no natural scale which could be responsible for quantum gravitational effects to become important. 
Typical bound states of elementary particles do not experience any new effects, since their gravitational interaction is
negligible. The conserved quantities in scattering processes are as usual the sums of the momenta of the asymptotically free states.

It is however also possible to start with a modification of the Lorentz-transformation for a 
free single particle and construct a {\sc QFT} based on this. For this, one assumes that the free
particle itself experiences the Planck mass as an upper bound on the energy scale.
Such a {\sc DSR} is a non-linear representation of the Lorentz-group 
\cite{Magueijo:2002am}, which can be cast in a form similar
to the above used by introducing variables that transform under the usual representation of the
Lorentz-group, ${\cal P}=(\epsilon,\pi)$, and a general map $F$ to the variables that then obey 
the new deformed transformation law $P=F({\cal P})$.  However, due to the different interpretation 
of the free particle's quantities, the 'pseudo-variable' ${\cal P}$ is not the one identified with 
the physical four momentum. Instead, energy and momentum of the in- and outgoing particles are 
identified with those that have the deformed transformation behavior.   
As it has turned out over the 
last years, such attempts result in serious conceptual problems, the most important 
being the question of which quantities are conserved in interactions and the soccer-ball problem, i.e. the 
question why the upper bound on the energy does not apply for many-particle states. 

Starting from an effective description in terms of a {\sc MDR} in the 
interaction region, we have shown in how far the here presented model is a reasonable candidate to describe 
strong gravitational effects in interactions. It has a clear interpretation 
for energy and momentum of the participating particles, and it does not suffer from the 
soccer-ball problem. As has been shown previously, the model has an ultraviolet regulator 
and  modified Feynman rules can be derived \cite{Hossenfelder:2004up}.  
 
From the present day status, it is not possible to decide which is the right description of
nature.  The model presented here is in certain regards more conservative and less spectacular than the usual
{\sc DSR} approach, 
but suffers from less conceptual problems.

{\bf{Apologies:}} Due to the minimal length of this article, I apologize for the lack of citations, equations
and paragraphs.

%%%%%%%%%%%%%%%%%%%%%%%%%%%%%%%%%%%%%%%%%%%%%%%%
%% BACKMATTER
%%%%%%%%%%%%%%%%%%%%%%%%%%%%%%%%%%%%%%%%%%%%%%%%

%%%%%%%%%%%%%%%%%%%%%%%%%%%%%%%%%%%%%%%%%%%
%% The following lines show an example how to produce a bibliography
%% without the help of the BibTeX program. This could be used instead
%% of the above.
%%%%%%%%%%%%%%%%%%%%%%%%%%%%%%%%%%%%%%%%%%%

\end{document}